\documentclass[12pt]{article}

\usepackage{graphicx} 
\usepackage{mathptmx} 
\usepackage{amsmath}
\usepackage{amssymb}
\usepackage{soul}
\usepackage{bm}
\usepackage[table]{xcolor}
\definecolor{lightgray}{gray}{0.9}
\usepackage{caption}
\usepackage{tabularray}
\usepackage{arydshln}
\UseTblrLibrary{booktabs}
\usepackage{natbib}
\usepackage{authblk}
\usepackage{tikz}
\usepackage{multirow}
\usepackage{makecell}
\usepackage{setspace}
\usepackage[title]{appendix}
\usepackage{chngcntr}     
\counterwithin{table}{section}   
\counterwithin{figure}{section}
\usetikzlibrary{shapes.geometric, arrows.meta, positioning}
\tikzstyle{box} = [
  rectangle, draw, rounded corners, 
  minimum width=1.5cm, inner ysep=1ex,
  text width=7cm, text depth=3ex, align=center
]
\tikzstyle{arrow} = [->, thick]
\usepackage[pdftex,linkcolor=black,pdfborder={0 0 0}]{hyperref} 
\usepackage[capitalise,nameinlink]{cleveref} 

\usepackage[a4paper, lmargin=0.08\paperwidth, rmargin=0.08\paperwidth, tmargin=0.08\paperheight, bmargin=0.08\paperheight]{geometry} 
\usepackage[normalem]{ulem}

\newcommand{\yy}{}
\newcommand{\bb}{}
\newcommand{\jj}{}
\newcommand{\note}[1]{}

\graphicspath{ {./img/} }

\title{Precision Dose-Finding Design for Phase I Oncology Trials by Integrating Pharmacology Data}

\author[1]{Kyong Ju Lee}
\author[2]{Yuan Ji}
\affil[1]{Department of Statistics, The University of Chicago}
\affil[2]{Department of Public Health Sciences, The University of Chicago}
\begin{document}
\pagenumbering{arabic}
\pagestyle{plain}
\maketitle

\begin{abstract}
      Phase I oncology trials aim to identify a safe dose—often the maximum tolerated dose (MTD)—for subsequent studies. Conventional designs focus on population-level toxicity modeling, with recent attention on leveraging pharmacokinetic (PK) data to improve dose selection. We propose the Precision Dose-Finding (PDF) design, a novel Bayesian phase I framework that integrates individual patient PK profiles into the dose-finding process. By incorporating patient-specific PK parameters (such as volume of distribution $V_i$ and elimination rate $k_i$), PDF models toxicity risk at the individual level, in contrast to traditional methods that ignore inter-patient variability. The trial is structured in two stages: an initial training stage to update model parameters using cohort-based dose escalation, and a subsequent test stage in which doses for new patients are chosen based on each patient's own PK-predicted toxicity probability. This two-stage approach enables truly personalized dose assignment while maintaining rigorous safety oversight. Extensive simulation studies demonstrate the feasibility of PDF and suggest that it provides improved safety and dosing precision relative to the continual reassessment method (CRM). The PDF design thus offers a refined dose-finding strategy that tailors the MTD to individual patients, aligning phase I trials with the ideals of precision medicine. \jj 
\end{abstract}

\textbf{Keywords}: Bayesian  design;    Dose response; \jj Pharmacodynamics; Pharmacokinetics;   Toxicity. \jj

\section{Introduction}
 
Phase I oncology trials are designed to evaluate drug safety and determine an appropriate dose for further testing, \yy like \jj 
the maximum tolerated dose (MTD)  \yy defined as  \bb the highest dose with a \jj toxicity rate not higher than a target rate. \jj  Dose-finding designs have traditionally fallen into two broad categories: simple algorithmic rule-based designs (such as the 3+3 and i3+3) and model-based designs that utilize statistical modeling of dose–toxicity relationships \citep{storer1989, Liu03032020, crm, mtpi, mtpi2, boin}. 
\yy Most designs \jj  assume that all patients share the same dose–toxicity \yy relationship, \jj ignoring individual heterogeneity in drug response.

In recognition of the limitations of the one-size-fits-all paradigm, efforts to incorporate patient-specific data—particularly pharmacokinetic (PK) measurements—into dose-finding began decades ago. For example, \citet{piantadosi1996} pioneered integrating PK information into a dose–toxicity model. Subsequent methods introduced summary PK exposure metrics, such as the area under the concentration–time curve (AUC) or maximum concentration ($C_{\max}$), as covariates in logistic or probit models for toxicity outcomes \citep{Whitehead08112007, ursino}. These approaches demonstrated that using PK data can \yy better capture the \jj  dose–toxicity relationships, even in small trials. In particular, \citet{toumazi2018} showed that incorporating PK measurements during dose escalation improved the estimation of the dose–toxicity curve without compromising the accuracy of MTD identification. More recently, semi-mechanistic designs have modeled the entire PK/PD time-course for each patient. Notably, the SDF design \citep{susdf} and the extended Bayesian semi‑mechanistic design of \citet{yang2024} used longitudinal PK and latent pharmacodynamic (PD) data to update toxicity probabilities, maintaining robust inference despite limited sample sizes. Building on \yy their work, \jj the PEDOOP design \citep{pedoop} separated the roles of PK and PD predictors for toxicity and efficacy, respectively, further bridging pharmacology and dose-finding. Each of these innovations represents a step toward more individualized dosing. 

Despite these advances, most current phase I designs continue to base dose decisions on population averages rather than individual patient characteristics. This can be problematic because patients vary widely in factors like organ function, comorbidities, and metabolism, which influence their ability to tolerate drugs \citep{kurzrock2023}. \yy The authors also note that \jj 
traditional flat dosing fails to account for such variability, making it challenging to \bb determine \jj ``the right dose for each patient". Recent research has underscored the benefits of accounting for patient heterogeneity. For instance, \citet{ollier2025} demonstrated that failing to include important patient covariates in a dose-finding model can drastically reduce the chance of selecting the correct dose and increase the risk of overdosing. Likewise, \citet{silva2023precisiondosefindingcancerclinical} proposed a precision dose-escalation design that identifies subgroups of patients with different MTDs when eligibility criteria are broadened. In their approach, patients are stratified into subpopulations \textit{a priori} or via sequential modeling, and separate MTDs are estimated for each subgroup. Such subpopulation-based designs represent progress toward personalized dosing, but they stop short of true N-of-1 dose adaptation since patients within each subgroup still receive the same recommended dose. Early work by \citet{babb2001} took the next step by deriving an individual MTD as a monotonic function of each patient's baseline anti-SEA antibody concentration, using the EWOC framework to control overdose probability. However, their method does not incorporate PK parameters. 

  In \yy this work, \jj  we introduce the Precision Dose-Finding (PDF) design, which achieves fully individualized dose selection by integrating each patient's own PK profile into the dose assignment process. The PDF design extends the PEDOOP framework one step further: rather than linking individual PK data to a population-level AUC metric, it uses each patient's personal PK parameters directly in the model. Specifically, we assume a one-compartment PK model for each patient with individual volume of distribution ($V_i$) and elimination rate ($k_i$), allowing us to compute that patient's drug exposure (AUC) and predict their toxicity risk in real time. By modeling toxicity probability as a function of patient-specific PK characteristics, the PDF method aligns with the paradigm of precision medicine and directly addresses the challenge of inter-patient variability in tolerance.

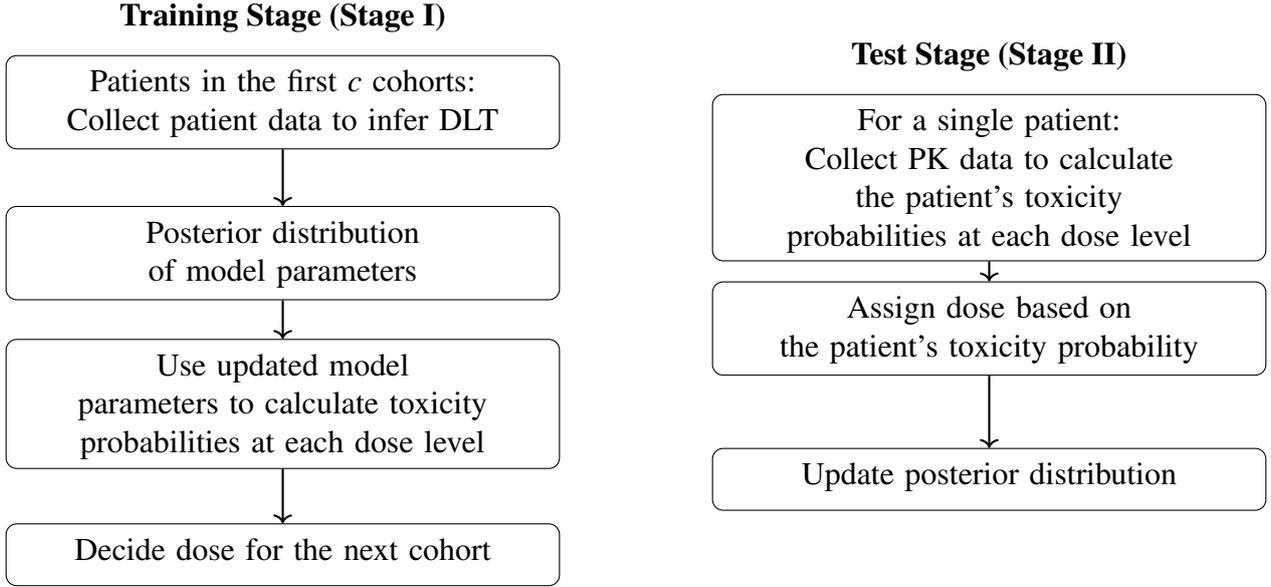
\begin{figure}[htbp]
\centering
\begin{tikzpicture}[node distance=2cm]
\node[box,  text depth=5.5ex] (c1) {Patients in the first $c$ cohorts: \\ Collect patient data to infer dose-limiting toxicity (DLT)};
\node[box, below of=c1] (post) {Posterior distribution of model parameters};
\node[box, below of=post, text depth=5.5ex] (dose) {Use updated model \\parameters to calculate toxicity\\ probabilities at each dose level};
\node[box, below of=dose, text depth=0.8ex] (escalate) {Decide dose for the next cohort};

\draw[arrow] (c1) -- (post);
\draw[arrow] (post) --(dose);
\draw[arrow] (dose) -- (escalate);

\node[box, right=of c1, yshift=-1cm, text depth=8.1ex] (c2) {For a single patient: \\Collect PK data to calculate \\\bb the patient's \jj toxicity \\probabilities at each dose level};
\node[box, below of=c2] (assign) {Assign dose based on \\the patient's toxicity probability};
\node[box, below of=assign, text depth=0.8ex] (update) {Update posterior distribution};

\draw[arrow] (c2) -- (assign);
\draw[arrow] (assign) -- (update);

\node[align=center] at (c1.north) [yshift=0.5cm] {\textbf{Training Stage (Stage I)}};
\node[align=center] at (c2.north) [yshift=0.5cm] {\textbf{Test Stage (Stage II)}};
\end{tikzpicture}

\caption{Schema of the PDF design. The training stage allocates patients using a conventional dose escalation method and focuses on updating model parameters. The test stage assigns doses to individual patients based on their personal toxicity probabilities derived from PK data.}
\label{flow}
\end{figure}

\yy PDF comprises  a training stage and \jj  a test stage (Figure \ref{flow}). During the training stage, patients are treated in cohorts using a conventional model-based escalation procedure to safely explore the dose range and iteratively update model parameters with accumulating toxicity and PK data. Once sufficient information has been gathered, the trial transitions to the test stage. In this stage, dose assignment becomes fully personalized: baseline characteristics of each new patient will be measured, and their PK parameters predicted.  Based on the predicted PK parameters, a predicted toxicity probability at each dose will be computed using the training data. The predicted MTD for that patient will then be administered \bb to \jj the patient.  The model is continually updated with each treated patient, and learning continues  in the test stage. This two-stage design ensures that the personalized dosing strategy is grounded in prior data and that its performance is validated in real time, thereby safeguarding patient safety while maximizing the precision of dose recommendations.


\jj 

We will begin with introducing the model assumptions of PDF in Section \ref{toxicmodel}, and explain how to allocate patients during the ``training" stage in Section \ref{doseescalationrules}. Next, Section \ref{mtdsel} will describe how the MTD of a trial can be determined, and Section \ref{precision} will describe how to perform precision dose-finding. In Section \ref{simstd}, we will present the simulation results and compare them with the \yy popular \jj  CRM design to evaluate the viability and performance of our approach.   In Section 4, we end the article with a brief discussion. \jj 

\section{Methods}

\subsection{  Probability Model \jj }\label{toxicmodel}
\yy For PK modeling, \jj  we assume a one-compartmental model for simplicity, which treats the human body as a single, uniform compartment and administers drugs via intravenous (IV) injection. Furthermore, we assume a first-order process to describe the rate of drug absorption and elimination \citep{Shargel2016-pk}.

Let $c_i(t)$ denote the drug concentration of patient $i \in \{1, ..., N\}$ at time $t$ after drug administration. The rate of a first-order process is $dc_i(t) / dt = -k_i c_i(t)$, with initial concentration $c_i(0) = d_i / V_i$. This yields the drug plasma concentration for patient $i$ at time $t$ as
\begin{align}
c_i(t | d_i, V_i, k_i) = \frac{d_i}{V_i} e^{-k_it}
\end{align}
where $d_i$ is \yy one of the $D$  \bb drug dosages \jj  administered to patient $i$, $d_i \in \{1, ..., D\}$, $k_i$ denotes the patient-specific elimination rate constant, and $V_i$ is \bb the \jj patient-specific volume of distribution. We also assume 
  $V_i>0$ and $k_i>0$ \jj follow   some prior distributions (e.g., gamma or log-normal) \jj with hyperparameters $\alpha$ and $\lambda$. 
  In our earlier work PEDOOP \citep{pedoop}, the priors for $V_i$ and $k_i$ can only take gamma distributions, a restriction we relax here. \jj 

Let $X_{ij}$ denote the observed   drug \jj concentration at time $t_j \in \{1, ..., T\}$ for patient $i$. We assume $X_{ij}$ follows a log normal distribution:
\begin{align}
\log(X_{ij}) &= \log\left(c_i(t_j \mid d_i, V_i, k_i)\right) + \epsilon_{ij}, \; \text{and }
\epsilon_{ij} \sim N(0, \sigma^2)
\end{align}
where $\epsilon_{ij}$ represents random error.

As described in PEDOOP, we compute the area under the concentration-time curve (AUC) for patient $i$ by integrating the drug concentration $c_i(t | d_i, V_i, k_i)$ from $0$ to $\infty$:
\begin{align}
    AUC_i(d_i, V_i, k_i) = \int_0^\infty c_i(t | d_i, V_i, k_i) dt = \frac{d_i}{V_i k_i}.
\end{align}

Next, we incorporate the AUC of patient $i$ as a covariate in modeling their toxicity probability.   Specifically, let $p_i$ denote the toxicity probability for patient $i$ and assume \jj 
\begin{align}\label{p_i}
p_i \triangleq p_i(V_i, k_i, \beta_0, \beta_1) &= logit^{-1}(\beta_0 + \beta_1 \times \log AUC_i(d_i, V_i, k_i)) \notag \\
& = logit^{-1}(\beta_0 + \beta_1 \times \log \frac{d_i}{V_i k_i}), 
\end{align}
  where $logit^{-1}(x) = 1/(1+e^{-x}).$ \jj 

Let $Y_i$ denote the binary DLT outcome for patient $i$   and $Y_i$ \jj  follows a Bernoulli distribution   given by \jj 
\begin{align}
Y_i | V_i, k_i, \beta_0, \beta_1 \sim Bernoulli(p_i(V_i, k_i, \beta_0, \beta_1)).
\end{align}
As a result, the joint posterior distribution of all parameters $\bm{\theta} = (V_i, k_i, \alpha_V, \lambda_V, \alpha_k, \lambda_k, \beta_0, \beta_1,  \sigma^2)$ given the  observed data $\mathcal{D}_I = \{(X_{ij}, Y_{i}, d_i)\}$, becomes
\begin{equation}
\begin{aligned}
     \pi(\bm{\theta} | \mathcal{D}_I) &\propto \prod_i^N \prod_j^T \frac{1}{\sigma \sqrt{2\pi}}\exp \left(-\frac{(\log(X_{ij})-\log\left(c_i(t_j \mid d_i, V_i, k_i)\right)\bb )^2 \jj}{2\sigma^2} \right) \\
    & \times \prod_{i}^N p_i^{Y_i} (1-p_i)^{1-Y_i} \\
    & \times \prod_i^N g(V_i | \alpha_V, \lambda_V) \times g(k_i | \alpha_k, \lambda_k) \notag \\
    & \times h(\beta_0) \times h(\beta_1) \notag \\ 
    & \times h(\alpha_V) \times h(\lambda_V) \times h(\alpha_k) \times h(\lambda_k) \times h(\sigma^2).
\end{aligned}
\end{equation}
Here, $g(\cdot)$ and $h(\cdot)$ denotes the corresponding prior distribution functions. We adopt the \yy hyper-prior \jj distributions  from PEDOOP, given by 
\begin{align}\label{priors}
    \alpha_V \sim Gamma(4, 1), \; \lambda_V \sim Gamma(1, 1) \notag \\
    \alpha_k \sim Gamma(3, 1), \; \bb \lambda_k \jj \sim Gamma(1, 1) \\
    \sigma \sim Gamma(3, 3), \; \beta_0 \sim N(-3, 100), \;\beta_1 \sim \log N(-1, 2). \notag
\end{align}
  These priors have been shown to lead to weak information on the $p_i$'s \citep{pedoop}. \jj Using the Markov chain Monte Carlo (MCMC) algorithm, we can generate posterior samples of $\bm{\theta} = (V_i, k_i, \alpha_V, \lambda_V, \\ \alpha_k, \lambda_k, \beta_0, \beta_1,  \sigma^2)$. 

\subsection{  Stage I: Cohort-Based Dose Escalation \jj }\label{doseescalationrules}
For the first $c$ cohorts, each with a fixed number of patients, we follow the conventional dose allocation rules, that is, assigning the same dose level to patients within the same cohort. Starting from the lowest dose level, we use the dose-level toxicity probability $\tilde{p}_d$ \yy (defined next) \jj to determine the appropriate dose for the next cohort via posterior predictive inference. 

Assume $n$ patients have been enrolled in the trial so far. \yy Let $\mathcal{D}_I$  represent \jj the cumulative data of stage I. \jj Let $Y_{new}$ denote the DLT outcome of a new patient in the next cohort treated at the dose level $Z_{new} = d$.
The predictive toxicity probability $\tilde{p}_d$ is computed as 
\begin{align}
   \tilde{p}_d & \triangleq Pr(Y_{new} =  1 | \mathcal{D}_I, Z_{new} = d) \notag \\
   & = \int Pr(Y_{new} = 1 | \bb V_{new}, k_{new}, \jj \beta_0, \beta_1, Z_{new} = d) \pi(\bb V_{new}, k_{new}, \jj \beta_0, \beta_1 | \mathcal{D}_I) \; d \bb V_{new} \jj \, d \bb k_{new} \jj \, d \beta_0 \, d \beta_1 \notag \\
   & = \int logit^{-1}(\beta_0 + \beta_1 \times \log \frac{d}{V_{new} k_{new}}) \pi(\bb V_{new}, k_{new}, \jj \beta_0, \beta_1 | \mathcal{D}_I) \; d \bb V_{new} \jj \, d \bb k_{new} \jj \, d \beta_0 \, d \beta_1 \\
   & \approx \frac{1}{M} \sum_{m=1}^M logit^{-1}(\beta_0^{(m)} + \beta_1^{(m)} log\frac{d}{V_{new}^{(m)} k_{new}^{(m)}}), \notag
\end{align}
  where notation $(m)$ denotes the $m$th sample from the MCMC simulation for the joint posterior $\pi(\bm{\theta} | \mathcal{D}_I)$, and  
$V_{new}^{(m)} = \sum_{i=1}^n \frac{V_i^{(m)}}{n}$ and $k_{new}^{(m)} =  \sum_{i=1}^n \frac{k_i^{(m)}}{n}$ for the new patients in the next cohort.  When the trial just starts, $\mathcal{D}_I$ is an empty set and $\pi(\bb V_{new}, k_{new}, \jj \beta_0, \beta_1 | \mathcal{D}_I)$ becomes the prior distribution of $\bb V_{new} \jj$ and $\bb k_{new} \jj$. \jj  

Following the CRM \citep{crm}, we choose the dose   $\tilde{d}$ for the next cohort where \jj 
\begin{align}\label{argmin}
    \tilde{d} = \arg\min_d |\tilde{p}_d - p_T|
\end{align}
and $p_T$ is a prespecified target toxicity rate (e.g., 0.3).

In addition, we \yy apply \jj  the following rules to enhance the efficiency and safety of the trial.
\begin{itemize}
    \item \textbf{Speed-up rule}: If no DLT outcomes are observed in any tested dose levels, escalate to the next higher dose for the next cohort to accelerate exploration. 
    
    \item \textbf{Safety rule 1 (no skip)}: If Equation (\ref{argmin}) suggests that the dose  should increase by two   or more \yy levels, \jj  we restrict escalation to only one level for the next cohort to address safety concerns \citep{Goodman1995}.
    
    \item \textbf{Safety rule 2 (dose exclusion \& early termination, \cite{tpi})}: 
    Let $n_d$ denote the number of patients treated at dose level $d$, and let $Y_d$ represent the number of patients   \bb who \jj experienced DLT at dose level $d$. Assume the prior distribution for the toxicity probability at dose $d$ is $p_d \sim Beta(0.05, 0.05)$, and the DLT outcome follows $Y_d | p_d \sim Bin(n_d, p_d)$. The posterior distribution of toxicity probability at dose $d$ is then $p_d | Y_d \sim Beta(Y_d + 0.05, n_d - Y_d + 0.05)$. A dose is regarded as safe if it satisfies 
    \begin{align}\label{safe2}
        Pr(p_d > p_T | Y_d, n_d) < s^*
    \end{align}
    where $p_T$ is the prespecified target toxicity rate and $s^*$ is a preselected safety threshold. If the suggested dose $\tilde{d}$ from (\ref{argmin}) is not safe, the algorithm instead recommends the highest dose that meets (\ref{safe2}) for the next cohort. If no dose meets this criterion, the trial will be terminated. Otherwise, the algorithm recommends the \yy highest \jj dose that satisfies both (\ref{argmin}) and (\ref{safe2}) for the next cohort.
    
    \item \textbf{Coherence principle}: At the end of the decision-making process, one more constraint is added to ensure escalation coherence \citep{cheung2005}. Suppose the current dose is $d$, and the recommended next dose based on the dose escalation rules described above is $d^*$. If the proportion of DLT outcomes at the current dose, i.e., $Y_d/n_d$, exceeds the target $p_T$ and $d^* > d$,  do not escalate; instead,    stay \jj at the current dose for the next cohort. Conversely, if $Y_d / n_d <p_T$ and $d^* < d$,  do not de-escalate, and \yy instead, \jj    stay \jj at the current dose level for the next cohort. 
\end{itemize}

\subsection{MTD Selection \yy for Stage I \jj }\label{mtdsel}
Once all patients in the first $c$ cohorts have been treated, we can select the MTD for the \yy training stage. \jj  To determine the MTD, we suggest comparing the posterior summary statistic of the dose-level toxicity probability $\hat{p}_d$ with the target toxicity probability $p_T$, and selecting the dose that minimizes the difference. As described in Safety rule 2, assuming a beta prior and a binomial distribution for DLT outcomes, the posterior mean of the toxicity probability at dose $d$ is: 
\[
\hat{p}_d= \frac{Y_d + 0.05}{n_d +0.1}
\]

Note that, to determine the MTD, $\hat{p}_d$ is transformed via isotonic regression to   ${\hat{p}^*}_d$ \jj using the pool adjacent violators algorithm \citep{robertson1988order}, which guarantees the monotonicity of the dose-toxicity relationship. 

Consequently, we select the dose as the MTD \yy for stage I \jj that satisfies 
\begin{align}
    d^{MTD} = \arg\min_d |  {\hat{p}^*}_d \jj  - p_T| 
\end{align}
from among the tested doses that are deemed safe \yy using \jj  criterion (\ref{safe2}).
Since   ${\hat{p}}_d^*$ \jj can be tied, we choose the MTD by the following rule: \\
(1) If $\hat{p}_{d^{MTD}} > p_T$, the MTD is the lowest dose among tied doses. \\
(2) If $\hat{p}_{d^{MTD}} \leq p_T$, the MTD is the highest dose among tied doses.

\subsection{Stage II: Precision Dose-Finding}\label{precision}

In precision dose-finding, each patient receives a predicted MTD using the posterior information learned from stage I. 
We recommend that 30\% of the total number of patients be treated under the precision dose-finding scheme, or that the training sample size be set to at least $21$, which is typical for phase I trials. Instead of following the dose escalation rules introduced in Section \ref{doseescalationrules} to determine dosage, we administer the dose based on each patient's individual toxicity profile. 

Specifically, for each patient, we estimate the predicted toxicity probability of patient $i$ if s/he is treated at dose $d$ by 
\begin{align}\label{p_ihat}
\hat{p}_{i,d}& = logit^{-1}(\hat{\beta}_0 + \hat{\beta}_1 \times \log \frac{d} {\yy \hat{V}_i \hat{k}_i \jj} ), 
\end{align}
where $\hat{\beta}_0$ and $\hat{\beta}_1$ are the posterior means of $\beta_0$ and $\beta_1$ from the first stage dose-finding, respectively. $(\hat{V}_i,\hat{k}_i)$ represent pre-dose predicted PK parameters for patient $i$. In precision dose-finding, we assume that individual PK parameters can be either directly measured or reliably predicted. In practice, one can obtain $(\hat{V}_i,\hat{k}_i)$ from a population PK (popPK) nonlinear mixed-effects model fit to Stage I PK data, combined with baseline covariates (e.g., body size and renal/hepatic function). Further details are provided in Section \ref{discussion}. 

In addition, the model parameters will continue to be updated as new data becomes available. \jj
Then patient $i$ is assigned to dose $d_i^*$ given by 
\begin{equation}\label{argmin2}
    d_i^*= \arg\min_d |  {\hat{p}_{i,d}} \jj  - p_T|.
\end{equation}
As a result, each new patient receives the dose that satisfies both \bb (\ref{argmin2}) \jj and (\ref{safe2}). 
This continues until the maximum number of patients for the trial is reached. Note that, as this precision dose-finding approach is implemented, selecting a single MTD for the overall trial becomes optional,   since each patient is associated with their own MTD $d_i^*$. \jj

\section{Simulation Study}\label{simstd}

\subsection{Simulation Setup}\label{simset}
We conducted \jj simulation studies of the PDF design under different scenarios and compared the results with 
the CRM design.
 Following the guidance of \citet{Goodman1995}, we performed \jj  1,000 trials across five distinct scenarios. Each scenario varied the true coefficient parameters $\beta_0$ and $\beta_1$, under the assumption that one of the dose levels is the true MTD. The true coefficient pairs used in these scenarios were $\{(-3, 1.5), (-4, 1.5), (-4.5, 1.5), (-2.5, 0.6), (-1, 1.2)\}$.

For $\bm{\theta} = (V_i, k_i, \alpha_V, \lambda_V, \alpha_k, \lambda_k, \beta_0, \beta_1,  \sigma^2)$, we assume the prior distributions   described \jj  in (\ref{priors}). In this simulation, $V_i$ and $k_i$   were \jj  assumed to follow gamma distributions, i.e., $
V_i \sim Gamma(4, 1), \; k_i \sim Gamma(3, 1), 
$ which is the same as PEDOOP. However, \jj later   in Section \ref{sensan1} \jj we  also conducted a sensitivity analysis using   different prior \jj distributions.

The target toxicity rate   was \jj  set to $p_T = 0.3$.   Simulated trials assumed \jj  five discrete dose levels $d \in \{15,30,60,90,120\}$ and   measured  \jj the drug concentration of each patient at time points $t_j \in \{1,3,5,7,12,\\24\}$ hours after injection. A total of 30 patients   \yy was \jj enrolled for each trial, with \jj the first 21 \bb receiving \jj doses using the traditional dose-finding algorithm \yy for stage I, \jj 
\yy and \jj the remaining 9 patients \yy following \jj  the precision dose-finding approach \yy for stage II. \jj 
We set $s^* = 0.95$ in (\ref{safe2}).

Note that due to the structure of our model, \yy the \jj true toxicity \yy probability \jj  at each dose level did not exist. Instead, we generated 2,000 random pairs of $V_i$ and $k_i$ from their prior distributions and plugged them into (\ref{p_i}), along with true $\beta_0$ and $\beta_1$, to simulate the true average toxicity \yy probability \jj for each dose level,   averaged across the 2,000 simulated patients. \jj These simulated averages were reported as \yy ``True avg $p_d$" \jj in the first line of each scenario \yy in Tables \ref{simres1to5} and \ref{last9tab}. \jj 

The CRM   design \yy was \jj  based on the model $p_d = \phi_d^{\exp(-\beta)}$,   $d=1, \ldots, 5,$ \jj where the skeleton $\phi = (\phi_1, ..., \phi_5)$ \yy was \jj  a set of prior guesses on toxicity probabilities at each dose level, and $\beta$ \yy followed \jj  a normal prior distribution with mean $0$ and standard deviation $2$. For fair comparison, the CRM   design \jj sample size \yy was \jj  set to $21$.

\subsection{Simulation Results}
The simulation results of our approach are presented in Tables \ref{simres1to5} and \ref{last9tab}, reporting the results of the first $21$ patients and the last $9$ patients, respectively. ``SC" denotes scenarios from $1$ to $5$; ``True avg $p_d$" represents the true average toxicity rates at each dose level; 
``$Y_d/n_d$" is the empirical toxicity rate at dose $d$; ``$n_d$" is the average number of patients assigned to each dose level over $1,000$ simulated trials; finally, ``Sel\%" is the proportion of simulated trials in which each dose is selected as the MTD. ``No MTD'' indicates early termination or failure to recommend a safe MTD.

\begin{table}[hbt!]
  \caption{Simulation results of the first 21 patients of scenarios 1 to 5. The column having a bold-faced number is the true MTD of that scenario.}
    \label{simres1to5}
  \centering
  \begin{tabular}{l|lccccccc}
             & & Dose levels & \textbf{1} & \textbf{2} & \textbf{3} & \textbf{4} & \textbf{5} & No MTD \\
    \toprule
    & & True avg $p_d$ & 0.151 & \textbf{0.287} & 0.470 & 0.586 & 0.665 & -- \\\cline{2-9}
    \multirow{5}{*}{SC 1} & \multirow{2}{*}{PDF} & $Y_d/n_d$ & 0.153 & 0.290 & 0.473 & 0.600 & 0.583 & -- \\
    & & $n_d$ & 6.255 & 8.808 & 5.028 & 0.717 & 0.024 & -- \\ 
    & & Sel\% & 0.160 & 0.540 & 0.257 & 0.032 & 0.001 & 0.010 \\ \cdashline{2-9}
    & \multirow{2}{*}{CRM} & $n_d $ & 7.698 & 7.821 & 4.677 & 0.756 & 0.048 & -- \\
    & & Sel\% & 0.158 & 0.503 & 0.299 & 0.038 & 0.002 & 0.000 \\

    \midrule
    & & True avg $p_d$ & 0.073 & 0.156 & \textbf{0.293} & 0.397 & 0.478 & -- \\\cline{2-9}
    \multirow{5}{*}{SC 2} & \multirow{2}{*}{PDF} & $Y_d/n_d$ & 0.066 & 0.154 & 0.300 & 0.400 & 0.458 & -- \\
    & & $n_d$ & 3.891 & 5.805 & 7.203 & 3.459 & 0.627 & -- \\ 
    & & Sel\% & 0.014 & 0.193 & 0.435 & 0.286 & 0.071 & 0.001 \\ \cdashline{2-9}
    & \multirow{2}{*}{CRM} & $n_d $ & 4.176 & 5.778 & 6.903 & 3.354 & 0.789 & -- \\
    & & Sel\% & 0.006 & 0.131 & 0.426 & 0.329 & 0.108 & 0.000 \\

    \midrule
    & & True avg $p_d$ & 0.049 & 0.110 & 0.220 & \textbf{0.310} & 0.385 & -- \\\cline{2-9}
    \multirow{5}{*}{SC 3} & \multirow{2}{*}{PDF} & $Y_d/n_d$ & 0.044 & 0.109 & 0.220 & 0.319 & 0.361 & -- \\
    & & $n_d$ & 3.516 & 4.659 & 6.447 & 4.746 & 1.632 & -- \\
    & & Sel\% & 0.004 & 0.055 & 0.298 & 0.399 & 0.228 & 0.016 \\ \cdashline{2-9}
    & \multirow{2}{*}{CRM} & $n_d $ & 3.642 & 4.617 & 6.345 & 4.455 & 1.941 & -- \\
    & & Sel\% & 0.000 & 0.040 & 0.248 & 0.405 & 0.307 & 0.000 \\    

    \midrule
    & & True avg $p_d$ & 0.111 & 0.158 & 0.218 & 0.260 & \textbf{0.293} & -- \\\cline{2-9}
    \multirow{5}{*}{SC 4} & \multirow{2}{*}{PDF} & $Y_d/n_d$ & 0.112 & 0.162 & 0.218 & 0.277 & 0.279 & -- \\
    & & $n_d$ & 4.671 & 5.445 & 5.472 & 3.819 & 1.539 & -- \\
    & & Sel\% & 0.019 & 0.114 & 0.291 & 0.324 & 0.249 & 0.003 \\ \cdashline{2-9}
    & \multirow{2}{*}{CRM} & $n_d $ & 4.896 & 5.361 & 5.634 & 3.432 & 1.677 & -- \\
    & & Sel\% & 0.012 & 0.107 & 0.262 & 0.315 & 0.304 & 0.000 \\       

    \midrule
    & & True avg $p_d$ & 0.422 & 0.593 & 0.747 & 0.819 & 0.860 & -- \\
    \cline{2-9}
    \multirow{5}{*}{SC 5} & \multirow{2}{*}{PDF} & $Y_d/n_d$ & 0.427 & 0.602 & 0.716 & 0 & 0 & -- \\
    & & $n_d$ & 13.146 & 2.643 & 0.162 & 0 & 0 & -- \\
    & & Sel\% & 0.494 & 0.031 & 0.002 & 0 & 0 & 0.473 \\ \cdashline{2-9}
    & \multirow{2}{*}{CRM} & $n_d $ & 19.131 & 1.713 & 0.156 & 0 & 0 & -- \\
    & & Sel\% & 0.952 & 0.047 & 0.001 & 0 & 0 & 0.000 \\       
    
  \end{tabular}
\end{table}

In scenarios $1$ and $2$, the true MTDs are dose levels $2$ and $3$, respectively. 
  The proposed PDF design \jj demonstrates strong performance in both scenarios, achieving the highest Sel\% at the correct   doses. \jj  Moreover, the highest $n_d$ values also occur at the correct MTDs, showing efficient patient allocation.   In addition, PDF allocates fewer patients and selects MTDs less frequently at   overly \jj toxic dose levels.

Both    PDF  and  CRM designs \jj assign the most patients to dose level $3$ in scenario $3$, where the true MTD is dose level $4$. However, the PDF design assigns a slightly higher average number of patients ($n_d$) to this dose level compared to the CRM. Additionally, both the PDF design and the CRM correctly identify the true MTD. Notably, our approach assigns fewer patients to, and selects less frequently, the   overly toxic \jj dose level $5$ compared to  CRM.

\begin{table}[hbt!]
  \caption{Simulation results of the last \yy nine \jj  patients of scenarios 1 to 5. \yy The column having a bold-faced number is the true MTD of that scenario. \jj}
    \label{last9tab}
  \centering
  \begin{tabular}{l|cccccc}
             & \text{Dose levels} & \textbf{1} & \textbf{2} & \textbf{3} & \textbf{4} & \textbf{5} \\ 
    \midrule
    & True avg $p_d$ & 0.151 & \textbf{0.287} & 0.470 & 0.586 & 0.665 \\\cline{2-7}
    SC 1 & $Y^*_d/n^*_d$ & 0.322 & 0.190 & 0.256 & 0.272 & 0.215 \\
    & $n^*_d$ & 2.668 & 3.297 & 1.458 & 0.581 & 0.901 \\

    \midrule
    & True avg $p_d$ & 0.073 & 0.156 & \textbf{0.293} & 0.397 & 0.478 \\\cline{2-7}
    SC 2 & $Y^*_d/n^*_d$ & 0.332 & 0.216 & 0.232 & 0.244 & 0.174 \\
    & $n^*_d$ & 0.834 & 1.836 & 1.999 & 1.302 & 3.020 \\

    \midrule
    & True avg $p_d$ & 0.049 & 0.110 & 0.220 & \textbf{0.310} & 0.385 \\\cline{2-7}
    SC 3 & $Y^*_d/n^*_d$ & 0.426 & 0.250 & 0.247 & 0.242 & 0.142 \\
    & $n^*_d$ & 0.646 & 1.062 & 1.548 & 1.266 & 4.478 \\

    \midrule
    & True avg $p_d$ & 0.111 & 0.158 & 0.218 & 0.260 & \textbf{0.293} \\\cline{2-7}
    SC 4 & $Y^*_d/n^*_d$ & 0.341 & 0.248 & 0.261 & 0.240 & 0.151 \\
    & $n^*_d$ & 0.615 & 0.940 & 1.103 & 1.048 & 5.260 \\

    \midrule
    & True avg $p_d$ & 0.422 & 0.593 & 0.747 & 0.819 & 0.860 \\
    \cline{2-7}
     SC 5 & $Y^*_d/n^*_d$ & 0.381 & 0.279 & 0.263 & 0.323 & 0.154 \\
    & $n^*_d$ & 3.718 & 0.502 & 0.133 & 0.031 & 0.026 \\
  \end{tabular}
\end{table}

  Scenario 4 presents a challenging case in which dose levels $3-5$ are all close to the target $p_T$. PDF and CRM \bb perform comparably, \jj allocating patients to mostly dose levels $3$ and $4$. With larger sample sizes, both designs may eventually recognize dose level $5$ as the true MTD.  
In scenario 5, all doses are \yy overly toxic. \jj PDF assigns slightly fewer patients to toxic doses with much \bb lower incorrect \jj MTD selection. \jj 


Table \ref{last9tab}   presents \jj the simulation results for the last nine patients, who underwent precision dose-finding.   Row 2 of each \bb scenario \jj presents $Y^*_d/n^*_d$, where $n^*_d$ and $Y^*_d$ are the numbers of patients among the last nine that are assigned to dose level $d$ and that experience DLT at dose level $d$, respectively. \jj  We can see that most $Y^*_d/n^*_d$ values are around \yy or smaller than \jj $p_T = 0.3$,   \bb regardless of \jj the true average $p_d$ values. This shows the benefit of precision dose-finding, since each patient is assigned based on his/her PK-implied optimal dose,  regardless \bb of what \jj the expected toxicity rate $p_d$ is. \jj 

In the test stage, each patient's dose is determined by his or her own PK profile. Because these PK values are predictions rather than directly observed quantities, they are subject to estimation error. To model this, we consider \yy adding errors to $V_i$ and $k_i$ and assume \jj 
\begin{align}
   \hat{V}_i = V_i + e_i, \quad e_i \sim \text{Truncated Normal}(0, \frac{\sigma_V^2}{9}, -V_i, \infty) \notag \\
   \hat{k}_i = k_i + g_i, \quad g_i \sim \text{Truncated Normal}(0, \frac{\sigma_k^2}{9}, -k_i, \infty) \notag
\end{align}
where each error term has mean $0$, variance equal to $1/9$ of the true variance, and is truncated below at the negative of its true value to prevent negative PK parameters. When dosing, we compute $\hat{p}_{i,d}$ by substituting $\hat{V}_i$ and $\hat{k}_i$ into the model. Table \ref{last9taberr} summarizes the simulation results under this framework. Although toxicity rates increased slightly, the overall pattern remains unchanged, \bb demonstrating \jj that the precision dose-finding stage is robust to realistic PK-prediction errors.

\begin{table}[hbt!]
  \caption{Simulation results of the last nine patients of scenarios 1 to 5 (with prediction errors of $V_i$ and $k_i$)}
    \label{last9taberr}
  \centering
  \begin{tabular}{l|cccccc}
             & \text{Dose levels} & \textbf{1} & \textbf{2} & \textbf{3} & \textbf{4} & \textbf{5} \\ 
    \midrule
    & True avg $p_d$ & 0.151 & \textbf{0.287} & 0.470 & 0.586 & 0.665 \\\cline{2-7}
    SC 1 & $Y^*_d/n^*_d$ & 0.375 & 0.204 & 0.246 & 0.286 & 0.213 \\
    & $n^*_d$ & 2.879 & 3.133 & 1.425 & 0.574 & 0.890 \\

    \midrule
    & True avg $p_d$ & 0.073 & 0.156 & \textbf{0.293} & 0.397 & 0.478 \\\cline{2-7}
    SC 2 & $Y^*_d/n^*_d$ & 0.408 & 0.226 & 0.219 & 0.249 & 0.181 \\
    & $n^*_d$ & 1.076 & 1.742 & 1.947 & 1.271 & 2.953 \\

    \midrule
    & True avg $p_d$ & 0.049 & 0.110 & 0.220 & \textbf{0.310} & 0.385 \\\cline{2-7}
    SC 3 & $Y^*_d/n^*_d$ & 0.426 & 0.250 & 0.247 & 0.242 & 0.142 \\
    & $n^*_d$ & 0.646 & 1.062 & 1.548 & 1.266 & 4.478 \\

    \midrule
    & True avg $p_d$ & 0.111 & 0.158 & 0.218 & 0.260 & \textbf{0.293} \\\cline{2-7}
    SC 4 & $Y^*_d/n^*_d$ & 0.387 & 0.271 & 0.271 & 0.234 & 0.162 \\
    & $n^*_d$ & 0.764 & 0.971 & 1.049 & 1.165 & 5.018 \\

    \midrule
    & True avg $p_d$ & 0.422 & 0.593 & 0.747 & 0.819 & 0.860 \\\cline{2-7}
    SC 5 & $Y^*_d/n^*_d$ & 0.408 & 0.238 & 0.362 & 0.167 & 0.259 \\
    & $n^*_d$ & 3.700 & 0.525 & 0.116 & 0.030 & 0.027 \\
  \end{tabular}
\end{table}

\begin{figure}
    \centering
    \includegraphics[width=0.9\linewidth]{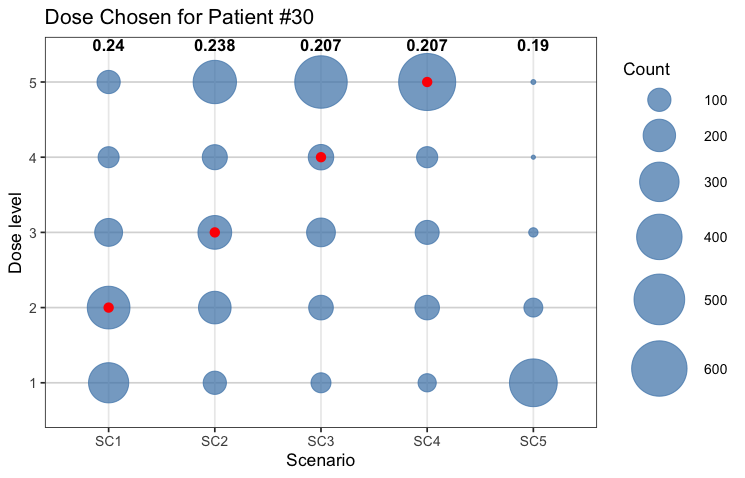}
    \caption{Dose assignment for the \yy last \jj patient treated in the precision dose-finding stage. Bubbles represent the number of \yy times patient 30 in each simulated trial is \jj  assigned to each dose level in each scenario. The numbers above the bubble columns report \bb the \jj proportion of DLTs patient 30 experienced in 1,000 trials in each scenario. 
    \jj  Red points denote the true MTD of the corresponding scenario.}
    \label{fig:bubble}
\end{figure}

Figure \ref{fig:bubble} displays the dose chosen for the \yy last  
patient (\# 30) \jj in each scenario. In general, the largest bubble (count) increases in SC 1 to 4, whereas SC5 keeps the patient at the lowest dose level. The DLT rates in each scenario are all below $p_T = 0.3$, implying that the precision-based dosing is safe.

\begin{figure}
   \centering
   \includegraphics[width=0.68\linewidth]{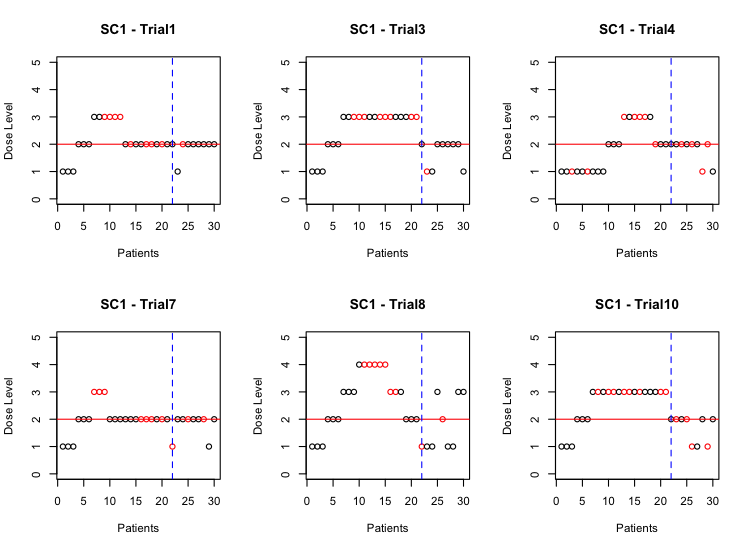}
   \caption{Dose assignments and DLT outcomes of six trials in scenario 1. Each point represents a patient. Black points: no DLT occurred; Red points: DLT occurred; Red horizontal lines: true MTD; Blue vertical dashed lines: start of precision dose-finding.}
   \label{doseassignprocess}
\end{figure}

We also \yy present outcomes of six randomly selected \jj trials from scenario 1 (Figure \ref{doseassignprocess}). It displays dose escalation patterns for the first $21$ patients and varying dose assignments for the last 9 patients, who participated in the precision dose-finding. Notably, the last $9$ patients tend to receive doses centered near the true MTD. The plots further reveal that some patients received the same dose but experienced different DLT outcomes. For instance, patient 22, 23, and 30 in ``SC1 - Trial3" had PK profiles $(V_i, k_i)$ of $(6.53, 3.18), (1.80, 1.65), \text{ and }(1.51, 1.34)$. Other trials, not shown here, also suggest that patients with a higher $V_i/k_i$ ratio may be more likely to experience DLT.

\begin{figure}[hbt!]
    \centering
    \includegraphics[width=0.9\linewidth]{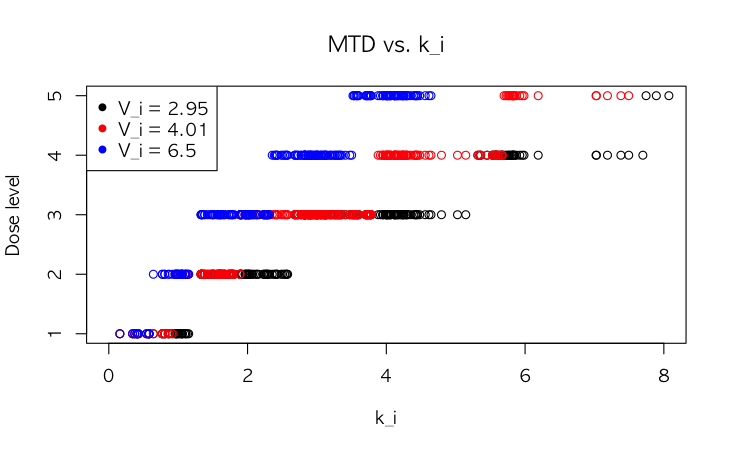}
    \caption{Examples of three \bb volumes \jj of distribution $V_i$ with varying elimination rate constants $k_i$ and their corresponding MTD. Each point represents a patient.}
    \label{mtdki}
\end{figure}

\yy To further illustrate the precision dose-finding, 
in Figure \ref{mtdki} we report the predicted MTD for a patient with different values of $V_i$ and $k_i$, given by 
$d_i = V_i k_i \exp(\frac{logit(p_i) - \hat{\beta}_0}{\hat{\beta}_1})$, rounded up to the nearest tested dose, where $p_i = 0.3$ and the coefficients are set to their posterior means. It can be seen \bb that as \jj $k_i$ increases, the predicted MTD also increases, implying a fast elimination rate helps increase tolerance to toxicity. Similarly, when $V_i$ increases, the predicted MTD also increases. For example, given $k_i=4$, the predicted MTD is dose level 3, 4, and 5 for $V_i$ value of 2.95, 4.01, and 6.5, respectively. This is also sensible if the toxicity is \bb inversely \jj related to $V_i$ as indicated in model \eqref{p_i}. \jj 

\subsection{Sensitivity Analysis}

\subsubsection{PK Parameter Priors}\label{sensan1}
As mentioned earlier, the choice of prior distributions for $V_i$ and $k_i$ is not restricted in our approach. Pharmacokinetic parameters such as clearance and volume of distribution are commonly modeled using log-normal distributions to ensure positivity and to capture the right-skewed nature of inter-individual variability \citep{mould2013}. In particular, the SDF design assumes that both $V_i$ and $k_i$ follow log-normal distributions, while the PEDOOP design assumes gamma distributions for computational convenience. Motivated by this standard practice, we additionally conduct a sensitivity analysis in which $V_i$ and $k_i$ are modeled using log-normal distributions instead of the gamma priors:
$$
V_i \sim \log N(\mu_V, \sigma^2_V), \; k_i \sim \log N(\mu_k, \sigma_k^2).
$$
We re-express the log-normal distribution parameters in terms of $\alpha$ and $\lambda$, while maintaining the same mean and variance as those used in the gamma distributions introduced in Section \ref{simset}.

\bb \cref{tab:app-sensres} and \cref{tab:app-senslast9} (See Appendix) \jj summarize the results of the sensitivity analysis compared to CRM. For the first $21$ patients, our approach performs comparably to CRM in the first three scenarios, reasonably assigning patients and accurately identifying the MTD. However, in scenario $4$, the PDF design selects a dose one level lower than the true MTD, while CRM successfully identifies it. In the final scenario, our approach results in a much lower Sel\% at toxic doses.

Regarding the precision dose-finding phase, the PDF design remains relatively safe across the first four scenarios and dose levels. However, in scenario $5$, the value of $Y_d^*/n_d^*$ is generally higher. This can be attributed to the change in the assumptions about the PK model.

\subsubsection{Stage I–Stage II Allocation Ratios}\label{sensan2}

The experiments above were conducted under a Stage I–Stage II allocation ratio of 7:3, although the proposed design readily accommodates other allocation ratios. To assess robustness with respect to this design choice, we conducted a sensitivity analysis under alternative allocation ratios of 5:5 and 6:4 in scenario 1, and compared the results with those of CRM under sample sizes of 15 and 18. The results are summarized in Appendix \cref{tab:simresdiffratioI} and \cref{tab:simresdiffratioII}. Across these settings, our method continues to outperform CRM in terms of operating characteristics, including assigning fewer patients to highly toxic doses and achieving higher accuracy in selecting the MTD, as well as selecting less toxic doses as the MTD. In addition, the observed toxicity rates, measured by $Y_d^*/n_d^*$, remain well controlled below the target level.
\section{Discussion}\label{discussion}

In this paper, we extend the PEDOOP \citep{pedoop} and the SDF designs \citep{susdf,yang2024} by proposing a Bayesian statistical model for the dose-toxicity relationship that simultaneously enables precision dose-finding. Our model inherits the strengths of the SDF designs, which \bb leverage \jj full PK profiles rather than relying on a summary PK measure, and incorporates the use of AUC as a covariate, as proposed in PEDOOP. However, unlike PEDOOP, which models toxicity probabilities at each dose level, our approach constructs a patient-specific model and uses posterior predictive probabilities to predict the toxicity risk for new patients in upcoming cohorts. Moreover, in contrast to PEDOOP, our model imposes no restrictions on the prior distributions of the volume of distribution and elimination rate constant, as long as they are clinically reasonable. A key innovation of our model is the new dose-finding framework, which assigns doses based on each patient's individual toxicity probability. Simulation results demonstrate that our approach is valid and safe, with low risk of overdosing. One potential benefit of this framework is that it enables early-phase trials to generate insights into PK-toxicity relationships, which may be informative for later-phase trials.

A prerequisite for the PDF design is an accurate \textit{pre‑dose} prediction of each patient's PK profile. A future direction is utilizing population pharmacokinetic (popPK) modeling for relating baseline covariates like body size, renal or hepatic function, and pharmacogenomic markers to key PK parameters. For example, by fitting a mixed‑effects model to the first 21 participants' rich PK samples, one can derive covariate effects and obtain patient-specific predictions for clearance $CL_i$ and volume $V_i$, from which $k_i$ can be computed via $k_i = CL_i/V_i$. Developing and validating such covariate-based popPK prediction models is beyond the scope of this paper. Instead, to reflect practical uncertainty in pre-dose PK prediction, we conducted simulations that introduce prediction errors into $(V_i,k_i)$ and evaluate the robustness of Stage II dose assignment (Table~3.3). Such model‑informed precision dosing has been advocated in oncology and therapeutic drug monitoring, and hybrid statistical/machine‑learning approaches have recently shown improved accuracy for antibiotics and targeted therapies \citep{iasonos2014, ma2024}. Alternatively, step-up (or intra-patient titration) dosing can serve as a potential solution. Under this approach, each Stage II patient first receives a conservative priming dose, followed by sparse PK sampling. These early PK measurements can then be used, together with appropriate pharmacokinetic models, to predict patient-specific PK parameter for the full dose and make subsequent dosing decisions. Deploying these predictions prospectively for the remaining patients enables one‑by‑one dose assignments while maintaining the Bayesian coherence of the PDF design.
\jj 

As in PEDOOP, we assume monotonicity of toxicity with respect to dose and use a first-order one-compartment model for IV injection. However, in practice, toxicity may not increase monotonically with dose and may instead follow a biphasic pattern, for example. Likewise, kinetic processes beyond first-order are possible. Thus, future work could explore extensions of our model that accommodate more flexible PK dynamics.

A potential limitation of our method is the computational burden, especially during the precision dose-finding process described in Section \ref{precision}, where MCMC sampling is used to update model parameters after each patient's enrollment. We report the running time of Stage I, Stage II, and the total procedure separately in Table~\ref{tab:comptime}.

Although our method does not define a single population-level MTD by combining data from Stage I and Stage II, it is designed to identify an individualized MTD for each patient. Such patient-specific dose recommendations can be particularly valuable for patients who require urgent treatment, where individualized risk–benefit considerations are critical. Moreover, if dose assignments in Stage II exhibit substantial variability across patients, clinicians may reasonably hesitate to recommend a single MTD for the entire population based solely on Stage I results, as this variability suggests meaningful heterogeneity in dose tolerance. In addition, as noted in PEDOOP, modern phase I trials are increasingly encouraged to identify the optimal biological dose (OBD), which motivates incorporating efficacy outcomes alongside toxicity. Given that toxicities may emerge late in treatment, identifying doses based solely on a toxicity endpoint may be inadequate \citep{tosi}. Therefore, future enhancements to the PDF design could be made from this perspective.

In Section \ref{precision}, we set the Stage I size to 70\% of the total number of patients, following a commonly used sample size in phase I trials. The sensitivity analysis indicates that our approach is robust to the choice of Stage I–Stage II allocation ratios. Nevertheless, we recommend setting the Stage I sample size larger than the number of dose levels, so that all investigational doses can be adequately explored before proceeding to Stage II. In addition, computational considerations may arise when the Stage II sample size becomes larger, since posterior updates are performed sequentially as each patient is enrolled. As an extension, we aim to explore the sample size determined by an iterative criterion -- e.g., measuring convergence in posterior summaries of model parameters -- to assess whether the trial is ready to enter the precision dose-finding stage. Alternatively, one could estimate the minimum training sample size required to support reliable inference. Otherwise, we recommend finding the desirable allocation ratio based on simulations using the PDF design.

\newpage

\bibliographystyle{apalike}
\bibliography{ref}

\newpage
\appendix

\section*{Appendix}                
\vspace{-0.6\baselineskip}   
\refstepcounter{section}            

\subsection{Sensitivity Analysis Results}\label{app:sens}
\begin{table}[hbt!]
  \caption{Simulation results of the first 21 patients of scenarios 1 to 5 under log-normal distributions.}
    \label{tab:app-sensres}
  \centering
  \begin{tabular}{l|lccccccc}
             & & Dose levels & \textbf{1} & \textbf{2} & \textbf{3} & \textbf{4} & \textbf{5} & No MTD \\
    \toprule
    & & True avg $p_d$ & 0.128 & \textbf{0.263} & 0.455 & 0.578 & 0.661 & -- \\\cline{2-9}
    \multirow{5}{*}{SC 1} & \multirow{2}{*}{PDF} & $Y_d/n_d$ & 0.139 & 0.269 & 0.434 & 0.607 & 0.544 & -- \\
    & & $n_d$ & 5.982 & 8.493 & 5.439 & 0.960 & 0.057 & -- \\ 
    & & Sel\% & 0.129 & 0.480 & 0.337 & 0.044 & 0.005 & 0.005 \\ \cdashline{2-9}
    & \multirow{2}{*}{CRM} & $n_d $ & 6.585 & 7.863 & 5.451 & 1.035 & 0.066 & -- \\
    & & Sel\% & 0.098 & 0.475 & 0.369 & 0.054 & 0.004 & 0.000 \\

    \midrule
    & & True avg $p_d$ & 0.056 & 0.132 & \textbf{0.269} & 0.378 & 0.463 & -- \\\cline{2-9}
    \multirow{5}{*}{SC 2} & \multirow{2}{*}{PDF} & $Y_d/n_d$ & 0.066 & 0.134 & 0.283 & 0.377 & 0.433 & -- \\
    & & $n_d$ & 3.846 & 5.274 & 7.074 & 3.909 & 0.897 & -- \\ 
    & & Sel\% & 0.010 & 0.141 & 0.410 & 0.331 & 0.108 & 0.000 \\ \cdashline{2-9}
    & \multirow{2}{*}{CRM} & $n_d $ & 3.843 & 5.214 & 6.960 & 3.969 & 1.014 & -- \\
    & & Sel\% & 0.001 & 0.095 & 0.356 & 0.400 & 0.148 & 0.000 \\

    \midrule
    & & True avg $p_d$ & 0.036 & 0.089 & 0.195 & \textbf{0.287} & 0.365 & -- \\\cline{2-9}
    \multirow{5}{*}{SC 3} & \multirow{2}{*}{PDF} & $Y_d/n_d$ & 0.039 & 0.092 & 0.210 & 0.288 & 0.354 & -- \\
    & & $n_d$ & 3.450 & 4.326 & 6.111 & 4.932 & 2.181 & -- \\
    & & Sel\% & 0.003 & 0.048 & 0.248 & 0.417 & 0.284 & 0.000 \\ \cdashline{2-9}
    & \multirow{2}{*}{CRM} & $n_d $ & 3.468 & 4.176 & 6.075 & 4.836 & 2.445 & -- \\
    & & Sel\% & 0.000 & 0.018 & 0.193 & 0.392 & 0.397 & 0.000 \\    

    \midrule
    & & True avg $p_d$ & 0.104 & 0.149 & 0.208 & 0.250 & \textbf{0.282} & -- \\\cline{2-9}
    \multirow{5}{*}{SC 4} & \multirow{2}{*}{PDF} & $Y_d/n_d$ & 0.118 & 0.147 & 0.217 & 0.254 & 0.287 & -- \\
    & & $n_d$ & 4.818 & 5.097 & 5.631 & 3.630 & 1.773 & -- \\
    & & Sel\% & 0.025 & 0.087 & 0.283 & 0.327 & 0.275 & 0.003 \\ \cdashline{2-9}
    & \multirow{2}{*}{CRM} & $n_d $ & 4.725 & 5.229 & 5.667 & 3.471 & 1.908 & -- \\
    & & Sel\% & 0.007 & 0.100 & 0.237 & 0.324 & 0.332 & 0.000 \\       

    \midrule
    & & True avg $p_d$ & 0.406 & 0.584 & 0.744 & 0.818 & 0.861 & -- \\\cline{2-9}
    \multirow{5}{*}{SC 5} & \multirow{2}{*}{PDF} & $Y_d/n_d$ & 0.414 & 0.592 & 0.710 & 0.444 & 0 & -- \\
    & & $n_d$ & 13.080 & 2.796 & 0.210 & 0.009 & 0 & -- \\
    & & Sel\% & 0.508 & 0.051 & 0.004 & 0.001 & 0 & 0.436 \\ \cdashline{2-9}
    & \multirow{2}{*}{CRM} & $n_d $ & 18.906 & 1.917 & 0.177 & 0 & 0 & -- \\
    & & Sel\% & 0.944 & 0.055 & 0.001 & 0 & 0 & 0.000 \\         
  \end{tabular}
\end{table}

\begin{table}[hbt!]
  \caption{Simulation results of the last nine patients of scenarios 1 to 5 under log-normal distributions.}
    \label{tab:app-senslast9}
  \centering
  \begin{tabular}{l|cccccc}
             & \text{Dose levels} & \textbf{1} & \textbf{2} & \textbf{3} & \textbf{4} & \textbf{5} \\ 
    \midrule
    & True avg $p_d$ & 0.128 & \textbf{0.263} & 0.455 & 0.578 & 0.661 \\\cline{2-7}
    SC 1 & $Y_d^*/n_d^*$ & \bb 0.320 \jj & 0.220 & 0.276 & 0.332 & 0.208 \\
    & $n_d^*$ & 2.359 & 3.055 & 1.647 & 0.651 & 1.203 \\

    \midrule
    & True avg $p_d$ & 0.056 & 0.132 & \textbf{0.269} & 0.378 & 0.463 \\\cline{2-7}
    SC 2 & $Y_d^*/n_d^*$ & 0.326 & 0.232 & 0.239 & 0.250 & 0.170 \\
    & $n_d^*$ & 0.680 & 1.639 & 1.904 & 1.367 & 3.397 \\

    \midrule
    & True avg $p_d$ & 0.036 & 0.089 & 0.195 & \textbf{0.287} & 0.365 \\\cline{2-7}
    SC 3 & $Y_d^*/n_d^*$ & 0.337 & 0.252 & 0.254 & 0.235 & 0.150 \\
    & $n_d^*$ & 0.371 & 0.927 & 1.322 & 1.309 & 5.067 \\

    \midrule
    & True avg $p_d$ & 0.104 & 0.149 & 0.208 & 0.250 & \textbf{0.282} \\\cline{2-7}
    SC 4 & $Y_d^*/n_d^*$ & 0.287 & 0.292 & 0.290 & 0.238 & 0.160 \\
    & $n_d^*$ & 0.537 & 0.888 & 1.137 & 1.120 & 5.289 \\

    \midrule
    & True avg $p_d$ & 0.406 & 0.584 & 0.744 & 0.818 & 0.861 \\\cline{2-7}
    SC 5 & $Y_d^*/n_d^*$ & 0.358 & 0.301 & 0.338 & 0.433 & 0.200 \\
    & $n_d^*$ & 3.780 & 0.655 & 0.142 & 0.030 & 0.035 \\
  \end{tabular}
\end{table}

\begin{table}[]
    \caption{Computation time (seconds) for Stage I, Stage II, and total runtime of a single simulated trial.}
    \label{tab:comptime}
    \centering
    \begin{tabular}{c|c|c|c|c}
         Scenario & Stage I (s) & Stage II (s) & Total (s) \\ \midrule
         SC 1 & 15.393 & 21.670 & 37.064 \\
         SC 2 & 15.115 & 23.295 & 38.410 \\
         SC 3 & 15.592 & 21.878 & 37.470 \\
         SC 4 & 15.217 & 22.014 & 37.230 \\
         SC 5 & 7.178 & 0.000 & 7.178
    \end{tabular}
\end{table}

\begin{table}[hbt!]
  \caption{Stage I simulation results for scenario 1 under varying Stage I–Stage II allocation ratios}
    \label{tab:simresdiffratioI}
  \centering
  \begin{tabular}{l|lcccccc}
             & & Dose levels & \textbf{1} & \textbf{2} & \textbf{3} & \textbf{4} & \textbf{5} \\
    \toprule
    & & True avg $p_d$ & 0.151 & \textbf{0.287} & 0.470 & 0.586 & 0.665 \\\cline{2-8}
    \multirow{5}{*}{\makecell{Stage I: 50\%\\ Stage II: 50\%}} & \multirow{2}{*}{PDF} & $Y_d/n_d$ & 0.151 & 0.287 & 0.478 & 0.609 & 0.333 \\
    & & $n_d$ & 5.589 & 6.156 & 2.847 & 0.297 & 0.003 \\ 
    & & Sel\% & 0.164 & 0.517 & 0.284 & 0.024 & 0.001 \\ \cdashline{2-8}
    & \multirow{2}{*}{CRM} & $n_d $ & 6.396 & 5.460 & 2.775 & 0.363 & 0.006 \\
    & & Sel\% & 0.204 & 0.412 & 0.279 & 0.089 & 0.016 \\

    \midrule
    & & True avg $p_d$ & 0.151 & \textbf{0.287} & 0.470 & 0.586 & 0.665 \\\cline{2-8}
    \multirow{5}{*}{\makecell{Stage I: 60\%\\ Stage II: 40\%}} & \multirow{2}{*}{PDF} & $Y_d/n_d$ & 0.150 & 0.290 & 0.479 & 0.608 & 0.417 \\
    & & $n_d$ & 5.922 & 7.476 & 3.942 & 0.510 & 0.012 \\ 
    & & Sel\% & 0.156 & 0.551 & 0.247 & 0.034 & 0.002 \\ \cdashline{2-8}
    & \multirow{2}{*}{CRM} & $n_d $ & 7.224 & 6.618 & 3.549 & 0.588 & 0.021 \\
    & & Sel\% & 0.190 & 0.458 & 0.302 & 0.045 & 0.005 \\
 
  \end{tabular}
\end{table}

\begin{table}[hbt!]
  \caption{Stage II simulation results for scenario 1 under varying Stage I–Stage II allocation ratios}
    \label{tab:simresdiffratioII}
  \centering
  \begin{tabular}{l|cccccc}
             & \text{Dose levels} & \textbf{1} & \textbf{2} & \textbf{3} & \textbf{4} & \textbf{5} \\ 
    \midrule
    & True avg $p_d$ & 0.151 & \textbf{0.287} & 0.470 & 0.586 & 0.665 \\\cline{2-7}
    \multirow{1}{*}{\makecell{Stage I: 50\%\\ Stage II: 50\%}} & $Y_d^*/n_d^*$ & \bb 0.342 \jj & 0.203 & 0.262 & 0.233 & 0.209 \\
    & $n_d^*$ & 4.225 & 4.567 & 2.269 & 1.202 & 2.513 \\

    \midrule
    & True avg $p_d$ & 0.151 & \textbf{0.287} & 0.470 & 0.586 & 0.665 \\\cline{2-7}
    \multirow{1}{*}{\makecell{Stage I: 60\%\\ Stage II: 40\%}} & $Y_d^*/n_d^*$ & 0.329 & 0.192 & 0.257 & 0.274 & 0.197 \\
    & $n_d^*$ & 3.511 & 4.069 & 1.937 & 0.764 & 1.557 \\
  \end{tabular}
\end{table}

\end{document}